\documentstyle[12pt,epsf,rotate]{article}

\newcommand{\n}{\hspace*{-2.5mm}}
\newcommand{\gsim}{\;\rlap{\lower 3.5 pt \hbox{$\mathchar \sim$}} \raise 1pt
 \hbox {$>$}\;}
\newcommand{\lsim}{\;\rlap{\lower 3.5 pt \hbox{$\mathchar \sim$}} \raise 1pt
 \hbox {$<$}\;}
\newcommand{\cl}{\mathop{{\mbox{Cl}}_2}\nolimits}
\newcommand{\li}{\mathop{{\mbox{Li}}_2}\nolimits}

\catcode`@=11
\newcount\@tempcntc
\def\@citex[#1]#2{\if@filesw\immediate\write\@auxout{\string\citation{#2}}\fi
  \@tempcnta\z@\@tempcntb\m@ne\def\@citea{}\@cite{\@for\@citeb:=#2\do
    {\@ifundefined
       {b@\@citeb}{\@citeo\@tempcntb\m@ne\@citea\def\@citea{,}{\bf ?}\@warning
       {Citation `\@citeb' on page \thepage \space undefined}}%
    {\setbox\z@\hbox{\global\@tempcntc0\csname b@\@citeb\endcsname\relax}%
     \ifnum\@tempcntc=\z@ \@citeo\@tempcntb\m@ne
       \@citea\def\@citea{,}\hbox{\csname b@\@citeb\endcsname}%
     \else
      \advance\@tempcntb\@ne
      \ifnum\@tempcntb=\@tempcntc
      \else\advance\@tempcntb\m@ne\@citeo
      \@tempcnta\@tempcntc\@tempcntb\@tempcntc\fi\fi}}\@citeo}{#1}}
\def\@citeo{\ifnum\@tempcnta>\@tempcntb\else\@citea\def\@citea{,}%
  \ifnum\@tempcnta=\@tempcntb\the\@tempcnta\else
   {\advance\@tempcnta\@ne\ifnum\@tempcnta=\@tempcntb \else \def\@citea{--}\fi
    \advance\@tempcnta\m@ne\the\@tempcnta\@citea\the\@tempcntb}\fi\fi}
\catcode`@=12
\begin{document}
\title{\vskip-3cm{\baselineskip14pt
\centerline{\normalsize\hfill LMU--04/96}
\centerline{\normalsize\hfill MPI/PhT/96--031}
\centerline{\normalsize\hfill MZ/Th--96--15}
\centerline{\normalsize\hfill TUM--HEP--247/96}
\centerline{\normalsize\hfill hep--ph/9606310}
\centerline{\normalsize\hfill May 1996}
}
\vskip1.5cm
Heavy-Higgs Lifetime at Two Loops}
\author{{\sc A. Frink$^1$, B.A. Kniehl$^{2,}$\thanks{Permanent address:
Max-Planck-Institut f\"ur Physik (Werner-Heisenberg-Institut),
F\"ohringer Ring~6, 80805 Munich, Germany.}, D. Kreimer$^1$, and
K. Riesselmann$^3$}\\
{\normalsize $^1$ Institut f\"ur Physik, Johannes-Gutenberg-Universit\"at,}\\
{\normalsize \phantom{$^1$} Staudinger Weg 7, 55099 Mainz, Germany}\\
{\normalsize $^2$ Institut f\"ur Theoretische Physik,
Ludwig-Maximilians-Universit\"at,}\\
{\normalsize \phantom{$^2$} Theresienstra\ss e~37, 80333 M\"unchen, Germany}\\
{\normalsize $^3$ Institut f\"ur Theoretische Physik,
Technische Universit\"at M\"unchen,}\\
{\normalsize \phantom{$^3$} James-Franck-Stra\ss e, 85747 Garching, Germany}}
\date{}
\maketitle
\begin{abstract}
The Standard-Model Higgs boson with mass $M_H\gg 2M_Z$ decays almost
exclusively to pairs of $W$ and $Z$ bosons.
We calculate the dominant two-loop corrections, of $O(G_F^2M_H^4)$, to the
partial widths of these decays.
In the on-mass-shell renormalization scheme, the correction factor is found to
be $1+14.6\%(M_H/\mbox{TeV})^2+16.9\%(M_H/\mbox{TeV})^4$, where the second term
is the one-loop correction.
We give full analytic results for all divergent two-loop Feynman diagrams.
A subset of finite two-loop vertex diagrams is computed to high precision
using numerical techniques.
We find agreement with a previous numerical analysis.
The above correction factor is also in line with a recent lattice calculation. 

\medskip
\noindent
PACS numbers: 12.15.Lk, 14.70.Fm, 14.70.Hp, 14.80.Bn
\end{abstract}

\newpage

\section{Introduction}

One of the longstanding questions of elementary particle physics is whether
nature makes use of the Higgs mechanism of spontaneous symmetry breaking to
endow the particles with their masses.
The Higgs boson, $H$, is the missing link sought to verify this theoretical
concept in the Standard Model (SM).
Its mass, $M_H$, is essentially a free parameter of the SM.
The failure of experiments at the CERN Large Electron-Positron Collider
(LEP~1) and the SLAC Linear Collider (SLC) to observe the decay
$Z\rightarrow f\bar f H$ has ruled out the mass range $M_H\le65.2$~GeV at
the 95\% confidence level \cite{gri}.
The discovery potential of LEP~2, with centre-of-mass (c.m.) energy
$\sqrt s=192$~GeV and luminosity $L=150$~pb$^{-1}$ per experiment, will reach
up to $M_H\approx95$~GeV \cite{gkw}.
We note in passing that $M_H$ lower bounds may also be derived theoretically
by demanding that the Higgs vacuum be (meta)stable \cite{lin}.

Several theoretical arguments bound $M_H$ from above.
The requirement that partial-wave unitarity in intermediate-boson scattering
at high energies be satisfied at tree level establishes an upper bound on $M_H$
at about $(8\pi\sqrt2/3G_F)^{1/2}\approx1$~TeV \cite{dic}.
This unitarity bound is significantly lowered by including one- \cite{joh}
and two-loop \cite{mah} corrections.
However, the improved bound depends on the energy scale, $\Lambda$, up to
which the SM is assumed to remain valid.
The same is true for the triviality bound, which is derived perturbatively
\cite{cab} by requiring that the running Higgs self-coupling, $\lambda(\mu)$,
stays finite for renormalization scales $\mu<\Lambda$.
Lattice computations, which take all orders of the perturbative expansion into 
account, yield a triviality bound of $M_H\approx700$~GeV \cite{das}.
The precise value of this bound depends on the specific form of the lattice
action and the regularization method used.

An alternative way of constraining $M_H$ from above is to require that the
Higgs sector be weakly interacting, so that perturbation theory is meaningful
\cite{vel}.
The resulting bounds depend somewhat on the considered process and the precise
definition of perturbation-theory breakdown, but they are independent of
assumptions concerning the scale $\Lambda$ beyond which new physics operates.
At one and two loops, the leading high-$M_H$ corrections to physical
observables related to Higgs-boson production or decay are of $O(G_FM_H^2)$
and $O(G_F^2M_H^4)$, respectively.
The one-loop corrections to the partial widths of the Higgs-boson decays to
pairs of fermions \cite{mve} and intermediate bosons \cite{mar} are relatively
modest, below 15\% at $M_H=1$~TeV if the on-mass-shell renormalization scheme
is employed.
The situation is similar for Higgs-boson production \cite{bak}.
For $M_H$ increasing, the $O(G_F^2M_H^4)$ corrections will eventually exceed
the $O(G_FM_H^2)$ ones in size.
In the case of the fermionic Higgs-boson decays, this happens at
$M_H\approx1.1$~TeV in the on-shell scheme \cite{dur,ghi}.
Obviously, the perturbative expansion in $G_FM_H^2$ ceases to usefully
converge for larger values of $M_H$.
The study of high-energy weak-boson scattering at two loops (with neglect of
trilinear couplings) leads to a perturbative upper bound on $M_H$ which
depends on the considered c.m.\ energy.
For $\sqrt s$ in the 1-TeV range, the $M_H$ upper bound comes out as low as
450~GeV~\cite{rie}.
However, this result may be relaxed using resummation techniques \cite{wil}.

Since a Higgs boson with $M_H\gg2M_Z$ decays almost exclusively to $W^+W^-$ and
$ZZ$ pairs, it is desirable, both from the theoretical and phenomenological
points of view, to find the two-loop $O(G_F^2M_H^4)$ corrections to the partial
widths of these decays as well.
This is the purpose of the present paper.
In contrast to the fermionic case \cite{dur,ghi}, this involves the
calculation of massive three-point functions at two loops.
Due to the presence of a large external momentum, heavy-mass expansion
techniques are not applicable here.
As in a previous analysis \cite{dur}, we take advantage of the
Goldstone-boson equivalence theorem \cite{cor} and neglect the gauge and
Yukawa couplings as well as the intermediate-boson masses.
In this approximation, the radiative corrections to the $H\to W^+W^-$ and
$H\to ZZ$ decay widths coincide.

In the evaluation of the relevant two-loop Feynman diagrams, both
infrared (IR) and ultraviolet (UV) divergences occur.
We calculate all divergent diagrams analytically, using dimensional
regularization.
This guarantees that the cancellation of the IR and UV divergences is manifest.
The remaining diagrams are individually devoid of singularities.
They are computed numerically in four space-time dimensions after
formulating the integrals in parallel and orthogonal space.
This has the advantage that potential numerical instabilities in connection
with the conventional Feynman parameterizations are avoided.
We list our results for the individual diagrams, so that they are available
to other authors for comparisons and further calculations.

We anticipate that we find good agreement with a recent, pioneering work by
Ghinculov \cite{agh}, which fully resorts to numerical methods.
In Ref.~\cite{agh}, the IR divergences are regularized by introducing a small
Goldstone-boson mass.
The final result is obtained by numerically taking the massless limit.
Another conceptual difficulty in Ref.~\cite{agh} is related to the treatment
of (endpoint) singularities in the Feynman-parameter integrals.
These are controlled by choosing suitable integration paths in the complex
plane with the aid of spline functions.
The values of the individual diagrams are not specified in Ref.~\cite{agh}.
We believe that, in view of the enormous complexity of the problem at hand, it
is indispensable to have at least two independent analyses that agree with
each other.

This paper is organized as follows.
In Sec.~2, we catalogue the contributing Feynman-graph topologies and explain
our renormalization framework.
In Sec.~3, we describe our analytical and numerical techniques.
In Sec.~4, we present our results for the individual two-loop vertex diagrams
and assemble the renormalized $H\to VV$ ($V=W,Z$) transition amplitude.
In Sec.~5, we explore the phenomenological implications of our result and
draw the conclusions.
The relevant one-loop three-point amplitudes to $O(\epsilon)$, where
$D=4-2\epsilon$ is the dimensionality of space time, are listed in the 
Appendix.

\section{Framework}

In this section, we set the stage for our calculation of the $H\to VV$
($V=W,Z$) decay widths to $O(G_F^2M_H^4)$.
As mentioned in the Introduction, this task may be greatly simplified in the
limit of interest, $M_H\gg2M_Z$, through the use of the Goldstone-boson
equivalence theorem \cite{cor}.
This theorem states that the leading high-$M_H$ electroweak contribution to a
Feynman diagram may be calculated by replacing the intermediate bosons
$W^\pm$, $Z$ with the would-be Goldstone bosons $w^\pm$, $z$ of the
symmetry-breaking sector of the theory.
In this limit, the gauge and Yukawa couplings may be neglected against the
Higgs self-coupling.
By the same token, the Goldstone bosons may be taken to be 
massless, and the fermion loops may be omitted.

Adopting the conventions of Ref.~\cite{pnm}, we may write the relevant
Lagrangian for the symmetry-breaking sector of the SM in terms of bare
quantities, marked by the subscript 0, as
\begin{eqnarray}
\label{eqsbs}
{\cal L}_0^{\rm SBS}&\n=\n&
\frac{1}{2}\partial_\mu{\bf w}_0\cdot\partial^\mu{\bf w}_0
+\frac{1}{2}\partial_\mu H_0\partial^\mu H_0
-\frac{1}{2}M_{w,0}^2 {\bf w}_0^2-\frac{1}{2}M_{H,0}^2 H_0^2\nonumber\\
&\n\n&-{\lambda_0\over4}\left({\bf w}_0^2+H_0^2\right)^2
-\lambda_0 v_0\left({\bf w}_0^2+H_0^2\right)H_0,
\end{eqnarray}
where $\lambda$ is the Higgs quartic coupling, $v$ is the Higgs vacuum 
expectation value, $H$ is the real scalar Higgs field, and the real scalar
SO(3) triplet, ${\bf w}=(w_1,w_2,w_3)$, is related to the Goldstone bosons,
$w^\pm$ and $z$, by $w^\pm=({w}_1\mp i{w}_2)/\sqrt{2}$ and $z={w}_3$,
respectively.
The tadpole counterterm, which cancels all tadpole contributions of
${\cal L}_0^{\rm SBS}$ order by order, has been omitted in writing
Eq.~(\ref{eqsbs}).
Consequently, all diagrams which include tadpole contributions need to be
dropped in calculations \cite{pnm}.

We work in the on-mass-shell renormalization scheme as formulated in
Refs.~\cite{dur,pnm}.
The residual SO(3) symmetry present in Eq.~(\ref{eqsbs}) is preserved in the
renormalization procedure, so that the $w^\pm$ and $z$ bosons share one mass
counterterm and one wave-function renormalization constant.
Specifically, the renormalization conditions are as follows:
(i) The mass counterterms, $\delta M_w^2$ and $\delta M_H^2$, are determined
so that the renormalized masses, $M_w^2=M_{w,0}^2-\delta M_w^2$ and
$M_H^2=M_{H,0}^2-\delta M_H^2$, coincide with the physical (pole) masses,
{\it i.e.}, the pole positions of the radiatively corrected propagators.
In the case of the Goldstone bosons, one has $M_w^2=0$.
(ii) The wave-function renormalization constants, $Z_w$ and $Z_H$, which enter
the  
relations $w_0^\pm=Z_w^{1/2}w^\pm$ and $H_0=Z_H^{1/2}H$ between the bare and 
renormalized fields are adjusted so that the on-shell propagators have unit
residues.
(iii) The physical Higgs quartic coupling and vacuum expectation value are 
fixed by requiring that $\lambda=G_FM_H^2/\sqrt2$ and $v=2^{-1/4}G_F^{-1/2}$,
respectively, where $G_F$ is Fermi's constant. 
This leads to the following relations \cite{dur,pnm}:
\begin{eqnarray}
\label{eqct}
\delta M_w^2&\n=\n&-\Pi_w^0(0),\nonumber\\
\delta M_H^2&\n=\n&-\mbox{Re}\Pi_H^0(M_H^2),\nonumber\\
\frac{1}{Z_w}&\n=\n&1-\frac{d}{dp^2}\Pi_w^0(p^2)\Bigr|_{p^2=0},\nonumber\\
\frac{1}{Z_H}&\n=\n&1-\frac{d}{dp^2}\mbox{Re}\Pi_H^0(p^2)\Bigr|_{p^2=M_H^2},
\nonumber\\
\lambda_0&\n=\n&\frac{\lambda}{Z_w}\left(1+
\frac{\delta M_H^2-\delta M_w^2}{M_H^2}\right),\nonumber\\
v_0&\n=\n&Z_w^{1/2}v,
\end{eqnarray}
where $\Pi_w^0(p^2)$ and $\Pi_H^0(p^2)$ are the self-energy functions of the
bare Goldstone and Higgs fields, respectively, calculated from
Lagrangian~(\ref{eqsbs}).
Two-loop expressions for $\delta M_w^2$, $\delta M_H^2$, $Z_w$, $Z_H$, and
$\lambda_0$ in terms of $M_H$ and $\lambda$ may be found in
Refs.~\cite{pnm,bij}.

The Feynman diagrams that contribute to the $H\to VV$ transition amplitude
through $O(G_F^2M_H^4)$ are depicted in Fig.~\ref{fig1}.
Dashed and solid lines represent Goldstone and Higgs bosons, respectively.
Adjacent propagators with identical four-momenta, which occur in connection 
with mass-counterterm insertions, are separated by solid circles.
The calligraphic labels stand for the values of the diagrams and are used in
our formulas;
the nomenclature extents the one introduced in Ref.~\cite{pnm}.
The underlying conventions are explained in Sec.~3 and the Appendix.

There are 6 one-loop diagrams ($\cal Q$, ${\cal B}_i$, and ${\cal C}_i$),
18 reducible two-loop diagrams (${\cal B}_i{\cal B}_j$, ${\cal B}_i{\cal C}_j$,
${\cal QT}_i$, and ${\cal QD}_i$), and
52 irreducible two-loop diagrams (${\cal A}_i$, ${\cal E}_i$, ${\cal L}_i$,
${\cal F}_i$, ${\cal G}_i$, ${\cal H}_i$, ${\cal I}_i$, ${\cal J}_i$, and
${\cal K}_i$).
In addition, there are 8 one-loop diagrams with one mass-counterterm insertion
($\delta M_a^2{\cal T}_i$ and $\delta M_a^2{\cal D}_i$, where $a=w,H$).
Further reducible two-loop diagrams emerge as products of one-loop vertex
and wave-function-renormalization diagrams; these are not counted here.
The one-loop diagrams need to be evaluated through $O(\epsilon)$.
The corresponding results for the tadpole ($\cal Q$) and two-point topologies 
(${\cal B}_i$ and ${\cal T}_i$) may be found in Ref.~\cite{pnm};
those for the three-point topologies (${\cal C}_i$ and ${\cal D}_i$) are
listed in the Appendix.
The irreducible two-loop two-point topologies (${\cal A}_i$, ${\cal E}_i$, and
${\cal L}_i$) may be taken from Ref.~\cite{pnm}.
Here, we regard two external legs that merge into the same vertex as one
point; solid circles are not counted as points.
For example, we view both ${\cal T}_i$ and ${\cal A}_i$ as two-point
topologies.

The main technical difficulty of the present paper is to tackle the
irreducible two-loop three-point topologies (${\cal F}_i$, ${\cal G}_i$,
${\cal H}_i$, ${\cal I}_i$, ${\cal J}_i$, and ${\cal K}_i$).
Topologies ${\cal F}_i$, ${\cal G}_i$, and ${\cal H}_i$ are ultraviolet 
divergent.
Topologies ${\cal H}_3$ and ${\cal H}_4$ are also plagued by infrared
singularities.
Topologies ${\cal I}_i$, ${\cal J}_i$, and ${\cal K}_i$ are UV and IR finite.
After describing our computational techniques in Sec.~3, we itemize the
results for topologies ${\cal F}_1$ through ${\cal K}_2$ in Sec.~4.

Calculating the combinatorial factors with the help of the software package
DIAGRAMMAR \cite{jmj}, we find the following master
formula for the two-loop $H\to VV$ transition amplitude in terms of
$\lambda_0$, $v_0$, and $M_H$:
\begin{eqnarray}
\label{eqmaster}
&\n\n&{\cal T}(H\to VV)=
-2\lambda_0v_0Z_wZ_H^{1/2}\left\{1
+i\lambda_0( 10{\cal B}_0 + 8{\cal B}_1 + 6{\cal B}_2)
+i(\lambda_0v_0)^2( 8{\cal C}_1 + 24{\cal C}_2 )
\phantom{\frac{1}{2}}\right.\nonumber\\
&\n\n&
{}-\lambda_0^2( 50 {\cal B}_0{\cal B}_0 + 36 {\cal B}_0{\cal B}_2 
                +16{\cal B}_1{\cal B}_1 + 18{\cal B}_2{\cal B}_2 
             +20{\cal Q}{\cal T}_0+8{\cal Q}{\cal T}_1+24{\cal Q}{\cal T}_2
                +36{\cal Q}{\cal T}_3
\nonumber\\
&\n\n&
\phantom{-\lambda^2(}
           {}+40{\cal A}_{10} +40{\cal A}_{1m} + 16{\cal A}_{20}
           +12{\cal A}_{2m}+24{\cal A}_3+36{\cal A}_4
\nonumber\\
&\n\n&
\phantom{-\lambda^2(}
           {}+60{\cal F}_1+4{\cal F}_2+24{\cal F}_3+40{\cal F}_4+16{\cal F}_5
              +24{\cal F}_6)
\nonumber\\
&\n\n&
{}-\lambda_0(\lambda_0 v_0)^2(
                          40{\cal B}_0{\cal C}_1+24{\cal B}_0{\cal C}_2
                         +32{\cal B}_1{\cal C}_1 + 96{\cal B}_1{\cal C}_2
                         +24{\cal B}_2{\cal C}_1+72{\cal B}_2{\cal C}_2
\nonumber\\
&\n\n&
\phantom{-\lambda_0(\lambda_0 v_0)^2(}
                      {}+16{\cal Q}{\cal D}_1+24{\cal Q}{\cal D}_{2a}
                         +24{\cal Q}{\cal D}_{2b}+144{\cal Q}{\cal D}_3
\nonumber\\
&\n\n&
\phantom{-\lambda_0(\lambda_0 v_0)^2(}
                      {}+80{\cal E}_1 +32{\cal E}_2+48{\cal E}_2^\star
                      +72{\cal E}_3 +144{\cal E}_4   +216{\cal E}_5
\nonumber\\
&\n\n&
\phantom{-\lambda_0(\lambda_0 v_0)^2(}
                      {}+40{\cal L}_1+32{\cal L}_{20}+144{\cal L}_{2m}
                      +96{\cal L}_3+216{\cal L}_5
\nonumber\\
&\n\n&
\phantom{-\lambda_0(\lambda_0 v_0)^2(}
                      {}+288{\cal G}_1+288{\cal G}_2+64{\cal G}_3+96{\cal G}_4
                         +48{\cal G}_5+80{\cal G}_6
\nonumber\\
&\n\n&
\phantom{-\lambda_0(\lambda_0 v_0)^2(}
                      {}+48{\cal I}_1+48{\cal I}_2+32{\cal I}_3+288{\cal I}_4
                         +160{\cal I}_5+96{\cal I}_6+32{\cal I}_7)
\nonumber\\
&\n\n&
{}-(\lambda_0 v_0)^4(864{\cal H}_1+288{\cal H}_2+64{\cal H}_3+96{\cal H}_4
                +144{\cal H}_5+48{\cal H}_6 +288{\cal K}_1+32{\cal K}_2
\nonumber\\
&\n\n&
\phantom{-(\lambda_0 v_0)^4(}
              {}+864{\cal J}_1+96{\cal J}_2+96{\cal J}_3+32{\cal J}_4
                +576{\cal J}_5+192{\cal J}_6+192{\cal J}_7+64{\cal J}_8)
\nonumber\\
&\n\n&
{}+\left[i\lambda_0(20{\cal T}_0+8{\cal T}_1)
+i(\lambda_0 v_0)^2(16{\cal D}_1+24{\cal D}_{2b})\right]\delta M_w^2
\nonumber\\
&\n\n&\left.
{}+\left[i\lambda_0(8{\cal T}_2+12{\cal T}_3)
+i(\lambda_0 v_0)^2(8{\cal D}_{2a}+48{\cal D}_{3})\right]\delta M_H^2
\right\}.
\end{eqnarray}
Here, the tree-level term has been factored out together with the wave-function
renormalization constants of the external legs;
in our two-loop approximation, these quantities need to be evaluated through
$O(\lambda^2)$.
The second and third terms within the curly bracket arise at one loop;
their prefactors of $\lambda_0$ and $v_0$ need only to be expanded through
$O(\lambda)$.
The residual terms are of genuine two-loop order;
here, $\lambda_0$ and $v_0$ may immediately be replaced with $\lambda$ and 
$v$,
and it is sufficient to insert the one-loop expressions for $\delta M_w^2$ and
$\delta M_H^2$.
Notice that we have already performed mass renormalization in
Eq.~(\ref{eqmaster}), so that $M_H$, which appears inside the expressions for
the various diagrams, is the only mass parameter left.

\section{Computational techniques}

As usual, we calculate the Feynman diagrams in $D=4-2\epsilon$ space-time
dimensions and introduce an unphysical 't~Hooft mass scale, $\mu$, to keep the
Higgs quartic coupling dimensionless.
After coupling renormalization, Eq.~(\ref{eqmaster}) will be finite and
$\mu$ independent in the physical limit $\epsilon\to0$.
To fix the notation, we write down the generic one-loop two-point
integral (with two scalar propagators):
\begin{equation}
\label{eqtwo}
{\cal B}(p^2,m_1^2,m_2^2)=\mu^{2\epsilon}\int\frac{d^Dq}{(2\pi)^D}\,
\frac{1}{(q^2-m_1^2+i\varepsilon)[(q+p)^2-m_2^2+i\varepsilon]}.
\end{equation}
The one-loop two-point diagrams in Fig.~\ref{fig1} are then defined as
\begin{eqnarray}
{\cal B}_0&\n=\n&{\cal B}(M_H^2,0,0),\nonumber\\
{\cal B}_1&\n=\n&{\cal B}(0,0,M_H^2),\nonumber\\
{\cal B}_2&\n=\n&{\cal B}(M_H^2,M_H^2,M_H^2),\nonumber\\
{\cal T}_0&\n=\n&\frac{\partial}{\partial m_1^2}{\cal B}(M_H^2,0,0),\nonumber\\
{\cal T}_1&\n=\n&\frac{\partial}{\partial m_1^2}{\cal B}(0,0,M_H^2),\nonumber\\
{\cal T}_2&\n=\n&\frac{\partial}{\partial m_2^2}{\cal B}(0,0,M_H^2),\nonumber\\
{\cal T}_3&\n=\n&\frac{\partial}{\partial m_1^2}{\cal B}(M_H^2,M_H^2,M_H^2).
\end{eqnarray}
Analytic results for these diagrams to $O(\epsilon)$ may be found in
Ref.~\cite{pnm}.
Note that, in contrast to Ref.~\cite{pnm}, we exclude the factors of $i$
connected with the scalar propagators in the definitions of the diagrams.
For the one-loop tadpole diagram, we have ${\cal Q}=M_H^2{\cal B}_1$.
The one-loop three-point diagrams, ${\cal C}_i$ and ${\cal D}_i$, are defined
and evaluated to $O(\epsilon)$ in the Appendix.

The two-loop diagrams are defined in analogy to Eqs.~(\ref{eqtwo}) and
(\ref{eqthree}).
We first consider the IR-finite two-loop diagrams in Fig.~\ref{fig1}.
In the limit $\epsilon\to0$, any member, $\cal M$, of this class may be cast
into the generic form
\begin{equation}
\label{eqtem}
{\cal M}=\frac{1}{(4\pi)^4(M_H^2)^{N-4}}
\left(\frac{4\pi\mu^2}{M_H^2e^\gamma}\right)^{2\epsilon}
\left[\frac{A}{\epsilon^2}+\frac{B}{\epsilon}+C+O(\epsilon)\right],
\end{equation}
where $\gamma$ is the Euler-Mascheroni constant, $N$ is the number of
propagators in ${\cal M}$, and $A$, $B$, and $C$ are finite, complex numbers.
The special form of the prefactor is to suppress the appearance of the familiar
term $\ln(4\pi)-\gamma$ in $B$ and $C$.
(By contrast, the $\gamma$ piece of the prefactor was expanded in
Ref.~\cite{pnm}.)
Coefficients $A$, $B$, and $C$ for diagrams ${\cal A}_i$, ${\cal E}_i$, and
${\cal L}_i$ are available in analytic form \cite{pnm}.
Diagrams ${\cal E}_1$ and ${\cal E}_2$ are IR divergent and will be discussed
further below.

The irreducible two-loop three-point diagrams with four propagators,
${\cal F}_i$, exhibit UV divergences in the form of quadratic and linear poles
in $\epsilon$, but they are IR safe.
They may be evaluated analytically from appropriate Feynman-parameter
representations, proceeding along the lines of Ref.~\cite{pnm}.
In the case of diagram ${\cal F}_1$, where all internal lines are massless,
we find agreement with Ref.~\cite{dav}.

Diagrams ${\cal G}_i$ and ${\cal H}_i$ are more involved, and it is
advantageous 
to apply dispersion relations in a way similar to Ref.~\cite{dis}.
Let us first concentrate on diagrams ${\cal G}_i$.
We observe that they contain a one-loop two-point subdiagram consisting of a
bubble and an adjacent propagator.
Calling the masses inside this bubble $m_1$ and $m_2$, the mass of the 
adjacent leg $m_3$, and the common four-momentum $q$, we may rewrite this
subdiagram as
\begin{equation}
\label{eqdr}
\frac{{\cal B}(q^2,m_1^2,m_2^2)}{q^2-m_3^2+i\varepsilon}
=\frac{{\cal B}(m_3^2,m_1^2,m_2^2)}{q^2-m_3^2+i\varepsilon}
-\frac{i}{(4\pi)^2}\int_{(m_1+m_2)^2}^\infty\frac{ds}{s}\,
\frac{\sqrt{\lambda(s,m_1^2,m_2^2)}}{s-m_3^2-i\varepsilon}\,
\frac{1}{q^2-s+i\varepsilon}+O(\epsilon),
\end{equation}
where $\lambda(s,m_1^2,m_2^2)=[s-(m_1+m_2)^2][s-(m_1-m_2)^2]$ is the
K\"all\'en function.
If Eq.~(\ref{eqdr}) is inserted in the expression for the one-loop seed
diagram, the first term turns into a product of the type
${\cal B}_i{\cal C}_j$, which contains all divergences.
In the second term, we may interchange the dispersion and loop integrations and
are left with a finite dispersion integral, which may be solved analytically.

Diagrams ${\cal H}_i$ may be treated in a similar fashion.
They contain a one-loop two-point subdiagram consisting of a
bubble and two identical adjacent propagators.
A useful representation of this subdiagram emerges by differentiating
Eq.~(\ref{eqdr}) with respect to $m_3^2$.
The divergences of diagrams ${\cal H}_i$ are contained in products of the
form ${\cal B}_i{\cal D}_j$.
In particular, this method allows us to separate the IR divergences of
diagrams ${\cal H}_3$ and ${\cal H}_4$.
By the way, this method may also be applied to the IR-divergent diagrams
${\cal E}_1$ and ${\cal E}_2$. 

Let us now turn to the cancellation of the IR divergences in
Eq.~(\ref{eqmaster}).
It is possible to extract the IR divergences along with the UV ones as poles
in $\epsilon$ using dimensional regularization.
In fact, this avenue was taken in Ref.~\cite{pnm}.
Here, we adopt an alternative strategy.
We identify a basic set of IR-divergent one-loop diagrams and show that their
prefactors vanish.
In this way, we do not need to solve any IR-divergent loop integral.
It is easy to see that the diagrams which emerge by duplicating any of the
Goldstone-boson propagators in diagrams ${\cal B}_0$, ${\cal B}_1$,
${\cal C}_1$, and ${\cal C}_2$ are IR divergent.
The resulting diagrams are ${\cal T}_0$, ${\cal T}_1$, ${\cal D}_1$, and
${\cal D}_{2b}$, respectively.
As for the irreducible two-loop diagrams, ${\cal E}_1$, ${\cal E}_2$,
${\cal H}_3$, and ${\cal H}_4$ are IR divergent.
Hence there are twelve IR-divergent terms in Eq.~(\ref{eqmaster}).
By means of the dispersion-relation method described above, we find that
${\cal E}_1-{\cal B}_1{\cal T}_0$, ${\cal E}_2-{\cal B}_1{\cal T}_1$,
${\cal H}_3-{\cal B}_1{\cal D}_1$, and ${\cal H}_4-{\cal B}_1{\cal D}_{2b}$
are finite.
Exploiting the one-loop identity \cite{pnm}
\begin{equation}
i\delta M_w^2=\lambda{\cal Q}+4(\lambda v)^2{\cal B}_1,
\end{equation}
it follows on that the twelve IR-divergent terms in Eq.~(\ref{eqmaster}) may be
grouped in four finite sets.
Specifically, we have
\begin{eqnarray}
\label{irfinsets}
(\lambda{\cal Q}-i\delta M_w^2){\cal T}_0+4(\lambda v)^2{\cal E}_1&\n=\n&
\lambda\left[-2\zeta(2)+1-i\pi+O(\epsilon)\right],\nonumber\\
(\lambda{\cal Q}-i\delta M_w^2){\cal T}_1+4(\lambda v)^2{\cal E}_2&\n=\n&
\lambda\left[-2+O(\epsilon)\right],\nonumber\\
(\lambda{\cal Q}-i\delta M_w^2){\cal D}_1+4(\lambda v)^2{\cal H}_3&\n=\n&
\frac{1}{v^2}\left[\frac{3}{8}\zeta(2)-\frac{1}{2}
+i\frac{\pi}{2}\ln2+O(\epsilon)\right],\nonumber\\
(\lambda{\cal Q}-i\delta M_w^2){\cal D}_{2b}+4(\lambda v)^2{\cal H}_4&\n=\n&
\frac{1}{v^2}\left[-\frac{\zeta(2)}{12}
-\frac{\sqrt3}{2}\cl\left(\frac{\pi}{3}\right)+\frac{\pi}{2\sqrt3}
+\frac{1}{2}+O(\epsilon)\right],
\end{eqnarray}
where $\zeta$ is Riemann's zeta function, with values
$\zeta(2)=\pi^2/6$ and $\zeta(3)\approx1.202\,056\,903$, and
$\cl{}$ is Clausen's integral, with value
$\cl(\pi/3)\approx1.014\,941\,606$.

In the remainder of this section, we describe the evaluation of the finite 
diagrams ${\cal I}_1$ though ${\cal K}_2$. 
This is achieved with the help of the methods developed
in Refs.~\cite{diss,2l2p,2l3p,CzKK}. These methods suggest a separation of the
integration space in an orthogonal and a parallel space.
The parallel space is defined to be the linear span of the exterior
momenta involved in the process under consideration.
For a three-point function, this is a two-dimensional space.
Each loop momentum is a sum of two vectors, one being
its projection onto the two-dimensional plane spanned by two
non-degenerate exterior momenta, the other one being orthogonal to
this plane. This splitting does not break Lorentz invariance \cite{MPLA}.
In the following, we define orthogonal- and parallel-space variables
for our purposes. We assume two independent exterior momenta,
$p_1$ and $p_2$, and two loop momenta, $l$ and $k$. Without loss of generality,
we assume $p_1$ and $p_2$ to be timelike.
Then, $e_1=p_1/\sqrt{p_1^2}$ and
$e_2=(p_2-p_2.e_1e_1)/\sqrt{(p_2.e_1)^2-p_2^2}$
are orthogonal unit vectors, with
$e_1^2=-e_2^2=1$ and $e_1.e_2=0$.
We further define
\begin{eqnarray}
l_0 &\n=\n& l.e_1,\nonumber\\
k_0 &\n=\n& k.e_1,\nonumber\\
l_1 &\n=\n&  l.e_2,\nonumber\\
k_1 &\n=\n&  k.e_2,\nonumber\\
l_\perp &\n=\n& l-l_0 e_1-l_1 e_2,\nonumber\\
k_\perp &\n=\n& k-k_0 e_1-k_1 e_2.
\end{eqnarray}
We then have
\begin{eqnarray}
e_i.l_\perp &\n=\n& 0,\nonumber\\
e_i.k_\perp &\n=\n& 0,\nonumber\\
l^2 &\n=\n& l_0^2-l_1^2-l_\perp^2,\nonumber\\
k^2 &\n=\n& k_0^2-k_1^2-k_\perp^2,\nonumber\\
l.p_1 &\n=\n& l_0 p_1.e_1,\nonumber\\
k.p_1 &\n=\n& k_0 p_1.e_1,\nonumber\\
l.p_2 &\n=\n& l_0 p_2.e_1-l_1 p_2.e_2,\nonumber\\
k.p_2 &\n=\n& k_0 p_2.e_1-k_1 p_2.e_2,\nonumber\\
l.k &\n=\n& l_0 k_0 -l_1 k_1 -l_\perp.k_\perp,\nonumber\\
(l+k)^2 &\n=\n& (l_0+k_0)^2 - (l_1 + k_1)^2 - l_\perp^2 -k_\perp^2
-2l_\perp.k_\perp.
\end{eqnarray}
Futhermore, we call the angle between the two-loop momenta in orthogonal
space
\begin{equation}
z  =  \frac{l_\perp.k_\perp}{\sqrt{l_\perp^2 k_\perp^2}}.
\end{equation}

In its most straightforward application, this approach delivers 
a two-fold integral representation for the two-loop two-point
function \cite{2l2p}. Typically, the remaining integrations are with
respect to the parallel-space component of each of the two loop
momenta.

For two-loop three-point functions \cite{2l3p}, and for
four-point functions as well \cite{2p4l}, this approach opens a route to
three-fold integral representations, regardless of the internal topology of
the function. In Ref.~\cite{CzKK}, this was improved to give a two-fold
integral representation for the planar three-point case.
Remarkably, the so-obtained integral representations are
solely defined on finite domains of integration \cite{CzKK}.
These results are sufficient to treat topologies ${\cal J}_i$
and, with slight modifications, topologies ${\cal I}_i$ as well.
A similar result for the three-point non-planar case was obtained recently
\cite{Frinkdipl}, and was used here for the first time to handle topologies
${\cal K}_i$.

We now wish to describe some features of the method.
Following separation into parallel-space and orthogonal-space variables,
the angular integration in the orthogonal space is complicated
due to the presence of propagators involving both loop momenta,
which results in a non-trivial integration over the variable $z$.
In fact, the $z$ integration delivers a result of the form
$f(P_i\mid_{l_\perp=k_\perp=0})$,
where $f$ is a function having a single branch-cut and $P_i$ is such
a propagator involving both loop momenta. In $f$, 
this propagator is evaluated at nullified
orthogonal loop momenta. Typically, in four-dimensional field theory,
$f$ turns out to be a logarithm in the two-loop two-point case,
and a square root in the two-loop three-point case.

By the very definition of a scalar field propagator, $P_i$ is a
quadratic form in parallel-space variables. Parallel-space integrations
over the real line are most conveniently performed with the help
of the residue theorem. To this end, we wish to close the contour
in either the upper or the lower halfplane. This is obscured by the
cut of $f(P_i\mid_{l_\perp=k_\perp=0})$, 
which is present in both halfplanes.
The problem is readily solved by using translations $l_0 \rightarrow 
l_0+l_1$ and $k_0 \rightarrow k_0 + k_1$. Using the $(+,-,-,-)$
signature of space-time, we see that these translations render
the propagators linear in the variables $l_1$ and $k_1$.
Accordingly, the cut defined by the equation
\begin{equation}
f(l_1,k_1)=f(P_i\mid_{l_\perp=k_\perp=0})=-c,
\end{equation}
with $c>0$, is now in either the lower or upper complex $l_1$ (resp.\ $k_1$)
plane.
This allows us to close the contour in the opposite halfplane.
The precise location of the cut becomes a function of the other integration
variables.
This results in constraint equations for the contributing residues, and thus
in constraints on the domains of integration for the remaining variables.
These domains turn out to be restricted to triangles in the
$(l_0,k_0)$ plane in the
case of two-loop three-point functions \cite{CzKK,Frinkdipl}.

At this stage, one is left with a four-fold integral representation.
There are two integrations for the remaining parallel-space variables
$l_0$ and $k_0$, and the moduli $s=l_\perp^2$ and $t=k_\perp^2$ still have to
be integrated over the positive real axis. These last two integrations can be 
carried out next.
These $s$ and $t$ integrals have the form \cite{CzKK,Frinkdipl} 
\begin{equation}
\int_0^\infty dt\,\frac{1}{t+t_0\pm i\eta}\int_0^\infty ds\,
\frac{1}{s+s_0(t)\pm i\eta}\,\frac{1}{\sqrt{(at+b+i\eta+cs)^2-4st}},
\end{equation}
with the cut of the square root chosen to be along the
positive real axis. The coefficients are functions of the parallel space
variables $l_0$ and $k_0$.
With due care to the analytic structure of the integrand,
one achieves the first integration by using either Landen
\cite{2l3p} or Euler transformations \cite{CzKK,Frinkdipl}.
The results allow one to interpret the $t$ integration as an integral
representation of dilogarithms and associated functions, and thus one
obtains a two-fold integral representation as the final result.
We once more stress that the final two integrations only cover finite
triangular domains. They are suitable for numerical integrations
using Gaussian quadrature.

This programme was first successfully applied to two-loop three-point
functions in Ref.~\cite{CzKK}, and then extended to the crossed topology in
Ref.~\cite{Frinkdipl}. For the crossed topology, the unavoidable presence
of two propagators involving both loop momenta results in a more
complicated cut structure. There are many more cases to be considered,
but the calculation still follows the scheme outlined above. 

\section{Results}

We now present our results for diagrams ${\cal F}_1$ through ${\cal K}_2$ in
the form of Eq.~(\ref{eqtem}).
The number of propagators is $N=4$ for ${\cal F}_i$,
$N=5$ for ${\cal G}_i$ and ${\cal I}_i$, and
$N=6$ for ${\cal H}_i$, ${\cal J}_i$, and ${\cal K}_i$.
Coefficients $A$, $B$, and $C$ are listed in Tables~\ref{tab1} and \ref{tab2}.
The formulas for coefficients $C$ of ${\cal F}_2$, ${\cal G}_5$, and
${\cal H}_5$ are somewhat lengthy, and we have introduced the following
constants:
\begin{eqnarray}
\label{eqcon}
{\rm F}_2&\n=\n&-\frac{4}{5}\zeta(3)
+\zeta(2)\left(-\frac{2}{5}\ln\frac{3-\sqrt5}{2}
-4\ln\frac{\sqrt5-1}{2}+\frac{6}{\sqrt5}+\frac{7}{2}\right)
+\frac{1}{12}\ln^3\frac{3-\sqrt5}{2}\nonumber\\
&\n\n&-\frac{2}{3}\ln^3\frac{\sqrt5-1}{2}
+\left(\sqrt5-1\right)\left(\frac{1}{2}\ln^2\frac{3-\sqrt5}{2}
-2\ln^2\frac{\sqrt5-1}{2}\right)-\frac{19}{2}\nonumber\\
&\n\approx\n&3.508\,941\,259,\nonumber\\
{\rm f}_2&\n=\n&4\ln^2\frac{\sqrt5-1}{2}
-2\sqrt5\ln\frac{\sqrt5-1}{2}-5\nonumber\\
&\n\approx\n&-1.921\,696\,013,\nonumber\\
{\rm G}_5&\n=\n&\int_0^{1/4}dx\,\frac{\sqrt{1-4x}}{x(1-x)}
\left[\li(-x)+\ln x\ln(1+x)\right]\nonumber\\
&\n\approx\n&-0.647\,466\,172,\nonumber\\
{\rm g}_5&\n=\n&-\int_0^{1/4}dx\,\frac{\sqrt{1-4x}}{x(1-x)}\,\ln(1+x)
\nonumber\\
&\n=\n&\sqrt3\left[\cl\left(\frac{2}{3}\pi + \arctan\sqrt{15}\right)
           + \cl\left(\frac{2}{3}\pi - \arctan\sqrt{15}\right)\right]
       -\frac{4}{\sqrt3}\cl\left(\frac{\pi}{3}\right)\nonumber\\
&\n\n&-2\ln^2\frac{\sqrt5-1}{2} + \frac{\pi}{\sqrt3}\ln2 \nonumber\\
&\n\approx\n&-0.177\,283\,264,\nonumber\\
{\rm H}_5&\n=\n&\int_0^{1/4}dx\,\frac{\sqrt{1-4x}}{(1-x)^2}
\left[\li(-x)+\ln x\ln(1+x)\right]\nonumber\\
&\n\approx\n&-0.065\,325\,931,\nonumber\\
{\rm h}_5&\n=\n&-\int_0^{1/4}dx\,\frac{\sqrt{1-4x}}{(1-x)^2}\ln(1+x)\nonumber\\
&\n=\n&-\frac{2}{\sqrt3}
\left[\cl\left(\frac{2}{3}\pi + \arctan\sqrt{15}\right)
           + \cl\left(\frac{2}{3}\pi - \arctan\sqrt{15}\right)\right]
       +\frac{8}{3\sqrt3}\cl\left(\frac{\pi}{3}\right)\nonumber\\
&\n \n& -\frac{\sqrt5}{2}\ln\frac{3-\sqrt5}{2} 
       -\frac{\pi}{\sqrt3}\left(\frac{2}{3}\ln2+\frac{1}{2}\right)
\nonumber\\
&\n\approx\n&-0.021\,441\,581.
\end{eqnarray}
The analytic results for ${\rm G}_5$ and ${\rm H}_5$ are rather messy, and we
refrain from writing them down here.

In Sec.~3, we discussed the elimination of the IR-divergent pieces in
Eq.~(\ref{eqmaster}).
Substituting in the remainder of Eq.~(\ref{eqmaster}) the expressions for
$\delta M_H^2$, $Z_w$, $Z_H$, $\lambda_0$, and $v_0$ in terms of $M_H$ and
$\lambda$ \cite{pnm} as well as the results for the various one- and two-loop 
diagrams, we may set $\epsilon=0$ and so obtain the physical result
\begin{eqnarray}
\label{eqfin}
&\n\n&{\cal T}(H\to VV)=
-2\lambda v\left\{1
+\hat\lambda\left[5\zeta(2) - 3\pi\sqrt3 + \frac{19}{2} + i\pi(2\ln2 - 5)
\right]\right.
\nonumber\\
&\n\n&
+\hat\lambda^2\left[
- 2\zeta(3) +\zeta(2)\left(- 45\pi\sqrt3+ 144\ln2 + \frac{1173}{2}\right) 
+ 4\left(33\sqrt3 + 4\pi\right)\cl\left(\frac{\pi}{3}\right) 
 \right.  
\nonumber\\
&\n\n&
\phantom{+\hat\lambda^2\left[\right.}
- 181\pi\sqrt3
- \frac{749}{8} 
+ 189{\rm L}_5
- 2{\rm F}_2 - 12{\rm G}_5 - 18{\rm H}_5
\nonumber\\
&\n\n&
\phantom{+\hat\lambda^2\left[\right.}
+ 12{\rm I}_1 + 12{\rm I}_2 + 8{\rm I}_3 + 72{\rm I}_4 
+ 40{\rm I}_5 + 24{\rm I}_6 + 8{\rm I}_7  
\nonumber\\
&\n\n&
\phantom{+\hat\lambda^2\left[\right.}
+ 108{\rm J}_1 + 12{\rm J}_2 + 12{\rm J}_3 + 4{\rm J}_4 
+ 72{\rm J}_5 + 24{\rm J}_6 + 24{\rm J}_7 + 8{\rm J}_8 
+ 36{\rm K}_1  + 4{\rm K}_2 
\nonumber\\
&\n\n&
\phantom{+\hat\lambda^2\left[\right.}
+ i\left( \zeta(2)\left( - 108\sqrt3\ln2 + 174\sqrt3 - 53\pi\right)
+ \pi\left(57\ln2 - \frac{41}{2}
- 2{\rm f}_2 - 12{\rm g}_5 - 18{\rm h}_5\right) \right.
\nonumber\\
&\n\n&
\phantom{+\hat\lambda^2\left[\right.}
+ 40{\rm i}_5 + 8{\rm i}_7 
+ 12{\rm j}_2  + 12{\rm j}_3 + 4{\rm j}_4 + 24{\rm j}_7 + 8{\rm j}_8 
+ 4{\rm k}_2
\left.\left.\left.\vphantom{\frac{1}{1}}\right)\right]\right\},
\end{eqnarray}
where $\hat\lambda=\lambda/(16\pi^2)=G_FM_H^2/(16\pi^2\sqrt2)$.
Here, ${\rm L}_5\approx0.923\,631\,827$ stems from the all-massive lemon
diagram, ${\cal L}_5={\rm L}_5/[(4\pi)^4M_H^2]+O(\epsilon)$, and is given in
terms of a two-dimensional integral in Eq.~(A86) of Ref.~\cite{pnm}, where it
is called ${\rm K}_5$.
All other constants are listed in Tables~\ref{tab1} and \ref{tab2} together
with Eq.~(\ref{eqcon}).
Inserting in Eq.~(\ref{eqfin}) the numerical values of these constants,
we find
\begin{eqnarray}
\label{eqnum}
{\cal T}(H\to VV)&\n\approx\n&
-2\lambda v\left\{1+\hat\lambda(1.400\,476 - 11.352\,791\,i)\right.
\nonumber\\
&\n\n&\left.-\hat\lambda^2[34.408\,2(43) + 21.003\,1(62)\,i]\right\}.
\end{eqnarray}
This agrees reasonably well with Eq.~(13) of Ref.~\cite{agh},
where the two-loop coefficient is found to be
$-[34.351(26)+20.999(20)\,i]$.
Taking the absolute squared of Eq.~(\ref{eqnum}), we obtain the correction
factor, $K_V$, in the relation $\Gamma(H\to VV)=K_V\Gamma_B(H\to VV)$ between
the radiatively corrected and Born values of the $H\to VV$ decay width as
\begin{eqnarray}
\label{eqkv}
K_V&\n\approx\n& 1 + 2.800\,952\,\hat\lambda
+ 62.030\,8(86)\,\hat\lambda^2
\nonumber\\
&\n\approx\n&1+14.629\%\left(\frac{M_H}{\mbox{1~TeV}}\right)^2
+16.921(2)\%\left(\frac{M_H}{\mbox{1~TeV}}\right)^4.
\end{eqnarray}

For comparison, we also list the analogous correction factor,
$K_f=\Gamma(H\to f\bar f)/\Gamma_B(H$
$\to f\bar f)$, for
the Higgs-boson decay to a pair of fermions \cite{dur,ghi}:
\begin{equation}
\label{eqkf}
K_f\approx1+11.058\%\left(\frac{M_H}{\mbox{1~TeV}}\right)^2
-8.908\%\left(\frac{M_H}{\mbox{1~TeV}}\right)^4.
\end{equation}
In the derivation of Eq.~(\ref{eqkf}), Yukawa couplings have been set to zero.
Therefore Eq.~(\ref{eqkf}) should be a useful approximation to the full
electroweak two-loop result as long as $M_f\ll M_H$.

\section{Discussion and conclusions}

We are now in a position to explore the phenomenological consequences of our 
result.
In Fig.~\ref{fig2}, we display the $H\to VV$ correction factor $K_V$ of
Eq.~(\ref{eqkv}) to $O(G_FM_H^2)$ and $O(G_F^2M_H^4)$ as a function of $M_H$.
For comparison, also the corresponding approximations for the $H\to f\bar f$
correction factor $K_f$ of Eq.~(\ref{eqkf}) are shown.
In both cases, the $O(G_FM_H^2)$ terms enhance the Born results.
This effect is slightly more pronounced for $K_V$.
In the case of $K_V$, the $O(G_F^2M_H^4)$ term acts in the same direction as
the $O(G_FM_H^2)$ term.
For $M_H$ increasing, it rapidly gains importance relative to the
$O(G_FM_H^2)$ term;
at $M_H\approx930$~GeV, both terms have the same numerical value, 12.6\%.
In the case of $K_f$, the $O(G_F^2M_H^4)$ term reduces and, for
$M_H\gsim1.114$~TeV, overcompensates the effect of the $O(G_FM_H^2)$ term.
Obviously, the perturbation expansions of $K_V$ and $K_f$ in $G_FM_H^2$ cease
to usefully converge for $M_H$ values in excess of about 1~TeV.
Consequently, this value may be considered as an upper bound on the Higgs-boson
pole mass $M_H$, if the SM is to be weakly interacting \cite{vel}.
We emphasize that this observation is independent of speculations concerning 
the energy scale up to which the SM is supposed to be valid.

The Goldstone-boson equivalence theorem \cite{cor} allowed us to extract power
corrections in $G_FM_H^2$.
In the high-$M_H$ limit, these are clearly dominant over corrections in
$G_FM_W^2$ and $G_FM_t^2$ arising from the gauge and Yukawa sectors of the SM,
respectively.
For the $H\to W^+W^-$ and $H\to ZZ$ decays, this was verified for the
$O(G_FM_H^2)$ terms in Refs.~\cite{hww,hzz}, where the respective partial
widths were completely calculated to one loop in the SM.
There is no obvious reason why the situation should be different at the
electroweak two-loop order.
QCD corrections enter the stage only at the two-loop level, in connection with
quark loops.
The dominant two- and three-loop QCD corrections to the $W^+W^-H$ and $ZZH$
couplings, of $O(\alpha_sG_FM_t^2)$ \cite{spi} and $O(\alpha_s^2G_FM_t^2)$
\cite{ste}, respectively, are well under control and suppressed against the
$G_FM_H^2$ power corrections for $M_H$ large.
The dominance of the $G_FM_H^2$ power corrections in the case of the $f\bar fH$
Yukawa couplings was discussed in great detail in Refs.~\cite{dur,dr}.
In this context, we should mention that the $O(G_FM_H^2)$ two-loop corrections
to the partial width of the loop-induced $H\to\gamma\gamma$ decay have been
computed \cite{koe} with the aid of the Goldstone-boson equivalence theorem.

Since a Higgs boson with mass $M_H\gg2M_Z$ decays almost exclusively to pairs
of intermediate bosons and top quarks, and we have gained control over the
dominant two-loop corrections to the respective partial widths, we are now
able to predict the lifetime of a high-mass Higgs boson with two-loop
precision.
For the reader's convenience, we list here the Born formulas of the relevant
partial decay widths:
\begin{eqnarray}
\label{eqdec}
\Gamma_B(H\to W^+W^-)&\n=\n&\frac{G_FM_H^3}{8\pi\sqrt2}(1-r_W)^{1/2}
\left(1-r_W+\frac{3}{4}r_W^2\right),\nonumber\\
\Gamma_B(H\to ZZ)&\n=\n&\frac{G_FM_H^3}{16\pi\sqrt2}(1-r_Z)^{1/2}
\left(1-r_Z+\frac{3}{4}r_Z^2\right),\nonumber\\
\Gamma_B\left(H\to t\bar t\,\right)&\n=\n&\frac{N_cG_FM_H^3}{16\pi\sqrt2}
r_t(1-r_t)^{3/2},
\end{eqnarray}
where $N_c=3$ and $r_a=4M_a^2/M_H^2$ ($a=W,Z,t$).
We compute the total Higgs-boson decay width as
\begin{equation}
\Gamma_H=K_V\left[\Gamma_B(H\to W^+W^-)+\Gamma_B(H\to ZZ)\right]
+K_f\Gamma_B\left(H\to t\bar t\,\right)\,.
\end{equation}
We display the result as a function of $M_H$ in Fig.~\ref{fig3}, assuming
$M_t=175$~GeV.
Comparing the Born, one-loop, and two-loop approximations,
the significance of radiative corrections to the heavy-Higgs lifetime,
$\tau_H=1/\Gamma_H$, becomes evident.

Equations~(\ref{eqkv}), (\ref{eqkf}), and (\ref{eqdec}) refer to the
on-mass-shell renormalization scheme and are thus independent of the unphysical
renormalization scale $\mu$, which enters via dimensional regularization.
By contrast, in other renormalization schemes such as the
$\overline{\mbox{MS}}$ scheme, there remains a weak $\mu$ dependence in the
loop-corrected expressions for the various Higgs-boson decay widths.
It is generally believed that the scheme and scale dependences of a
calculation up to a given order indicate the size of the unknown higher-order
contributions, {\it i.e.}, they provide us with an estimate of the theoretical
uncertainty.
Applying this empirical rule to the fermionic and bosonic Higgs-boson decays,
one may conclude that the respective perturbative results are likely to 
become unreliable already for $M_H\gsim700$~GeV~\cite{boc,nie}.
In fact, for $M_H\gsim700$~GeV, the $\mu$ dependences of the
$\overline{\mbox{MS}}$ results are no longer reduced if the two-loop
corrections are included \cite{nie}.

Finally, we should mention that Monte-Carlo studies on finite lattices
provide us with a hint on the all-order relation between the mass and the
bosonic decay width of the Higgs boson \cite{goe,rum}.
In Ref.~\cite{goe}, elastic $\pi\pi$ scattering was analyzed nonperturbatively
in the framework of the four-dimensional O(4)-symmetric nonlinear $\sigma$
model in the broken phase, and the $\sigma$ resonance was observed.
The mass $m_\sigma$ and width $\Gamma_\sigma$ of this resonance could be
extracted from the measured scattering phases.
At the same time, the pion mass $m_\pi$ and wave-function-renormalization
constant $Z$ as well as the field expectation value $\Sigma$ in infinite 
volume were determined.
In this way, it is possible to compare the nonperturbative value of
$\Gamma_\sigma$ with the respective perturbative value obtained by inserting
$m_\pi$, $m_\sigma$, $Z$, and $\Sigma$ into the tree-level formula for
$\Gamma_\sigma$, given by Eqs.~(9.4) and (9.5) of Ref.~\cite{goe}.
For the central simulation point, the nonperturbative result is found to
exceed the tree-level one by approximately 16.7\% \cite{wes}.
It is interesting to confront this all-order correction with the one- and
two-loop corrections of Eq.~(\ref{eqkv}).
If we identify $m_\sigma$, $Z$, and $\Sigma$ of Ref.~\cite{goe} with our
parameters $M_H$, $Z_w$, and $v_0$, respectively, we have
$2\lambda=(M_H/v)^2=Z(m_\sigma/\Sigma)^2$, where the lattice constant cancels
on the right-hand side.
We so find that the central simulation point of Ref.~\cite{goe} corresponds to
$M_H\approx727$~GeV, for which the corrections through $O(G_FM_H^2)$ and
$O(G_F^2M_H^4)$ in Eq.~(\ref{eqkv}) amount to approximately 7.7\% and 12.5\%,
respectively.
The tree-level, one-loop, two-loop, and nonperturbative values of $K_V$ at
this value of $M_H$ are indicated by the crosses in Fig.~\ref{fig2}. 
In other words, the one- and two-loop terms of $K_V$ are 7.7\% and 4.7\%, 
respectively, while the higher-order terms apparently add up to a total of
4.3\%.
This is a very reasonable result, which gives support to the notion that the
perturbative expansion in $G_FM_H^2$ of the bosonic Higgs-decay width is still
usefully convergent for this value of $M_H$, in accordance with the 
conclusions drawn above on the basis of Fig.~\ref{fig2}.
However, this observation should be taken with a grain of salt.
The analysis of Ref.~\cite{goe} was performed with $m_\pi$ finite, while we
set $M_w=0$ for the Goldstone-boson pole mass.
Nevertheless, we believe that such finite-mass effects will not drastically
change our conclusions as long as we compare the relative corrections to the
decay widths rather than the absolute values of the latter.
This assumption is substantiated by the detailed inspection of the full
one-loop results for the $H\to W^+W^-$ and $H\to ZZ$ partial decay widths
\cite{hww,hzz}.

\bigskip
\centerline{\bf ACKNOWLEDGEMENTS}
\smallskip\noindent
We thank A. Ghinculov for providing us with some partial results of
Ref.~\cite{agh}, and M. G\"ockeler, M. L\"uscher, P. Weisz, and J. Westphalen
for valuable advice regarding the comparison of our result with
Ref.~\cite{goe}.
The collaboration on this project was initiated during the Workshop on
{\it Higher Order Perturbative Corrections in the Standard Model} at the
{\it Aspen Center for Physics}.

\begin{appendix}
\section{Appendix: One-loop three-point diagrams}

Here, we present the relevant one-loop three-point diagrams to $O(\epsilon)$.
We write the generic one-loop three-point integral as
\begin{eqnarray}
\label{eqthree}
&\n\n&{\cal C}(p^2,k^2,(p+k)^2,m_1^2,m_2^2,m_3^2)\nonumber\\
&\n\n&=\mu^{2\epsilon}\int\frac{d^Dq}{(2\pi)^D}\,
\frac{1}{(q^2-m_1^2+i\varepsilon)[(q+p)^2-m_2^2+i\varepsilon]
[(q+p+k)^2-m_3^2+i\varepsilon]}.
\end{eqnarray}
The one-loop three-point diagrams in Fig.~\ref{fig1} are then defined as
\begin{eqnarray}
{\cal C}_1&\n=\n&{\cal C}(0,0,M_H^2,0,M_H^2,0),\nonumber\\
{\cal C}_2&\n=\n&{\cal C}(0,0,M_H^2,M_H^2,0,M_H^2),\nonumber\\
{\cal D}_1&\n=\n&\frac{\partial}{\partial m_1^2}
{\cal C}(0,0,M_H^2,0,M_H^2,0),\nonumber\\
{\cal D}_{2a}&\n=\n&\frac{\partial}{\partial m_2^2}
{\cal C}(0,0,M_H^2,0,M_H^2,0),\nonumber\\
{\cal D}_{2b}&\n=\n&\frac{\partial}{\partial m_2^2}
{\cal C}(0,0,M_H^2,M_H^2,0,M_H^2),\nonumber\\
{\cal D}_3&\n=\n&\frac{\partial}{\partial m_1^2}
{\cal C}(0,0,M_H^2,M_H^2,0,M_H^2).
\end{eqnarray}
We find
\begin{eqnarray}
{\cal C}_1&\n=\n&
   \frac{i}{(4\pi)^2M_H^2}
\left(\frac{4\pi\mu^2}{M_H^2e^\gamma}\right)^\epsilon
\left\{
   -\frac{\zeta(2)}{2}-i\pi\ln2 \right. \nonumber\\
&\n\n&
   \left.+\epsilon\left[-\frac{19}{8}\zeta(3)+\frac{9}{2}\zeta(2)\ln2 
   +i\pi\left(-\zeta(2)+\frac{\ln^22}{2}\right)\right]
   + O(\epsilon^2)\right\},
   \nonumber\\
{\cal C}_2&\n=\n&
   \frac{i}{(4\pi)^2M_H^2}
\left(\frac{4\pi\mu^2}{M_H^2e^\gamma}\right)^\epsilon
\left\{
   -\frac{2}{3}\zeta(2)
   +\epsilon\left[\frac{2}{9}\zeta(3)-\frac{4\pi}{9}\cl
\left(\frac{\pi}{3}\right)\right] 
   + O(\epsilon^2)\right\},
   \nonumber\\
{\cal D}_{2a}&\n=\n& 
   \frac{i}{(4\pi)^2M_H^4}
\left(\frac{4\pi\mu^2}{M_H^2e^\gamma}\right)^\epsilon
\left\{
   i\frac{\pi}{2}
   +\epsilon\left[-\frac{5}{4}\zeta(2)+i\frac{3\pi}{2}\ln2\right]
   + O(\epsilon^2)\right\},
   \nonumber\\
{\cal D}_3&\n=\n&
   \frac{i}{(4\pi)^2M_H^4}
\left(\frac{4\pi\mu^2}{M_H^2e^\gamma}\right)^\epsilon
\left\{\frac{\pi}{3\sqrt3}
   +\epsilon\left[\frac{2}{3}\zeta(2)+\frac{4}{3\sqrt3}\cl
\left(\frac{\pi}{3}\right) 
   -\frac{\pi}{3\sqrt3}\ln3\right]
   + O(\epsilon^2)\right\}.\nonumber\\
\end{eqnarray}
As already stated in Sec.~3, ${\cal D}_1$ and ${\cal D}_{2b}$ are IR 
divergent.
We need not calculate them, since their prefactors in Eq.~(\ref{eqmaster}) 
vanish according to Eq.~(\ref{irfinsets}).

\end{appendix}

\newpage

\renewcommand{\arraystretch}{1.5}
\begin{table}[p]

\centerline{\bf TABLES}

\caption{Coefficients $A$, $B$, and $C$ to be inserted in
template~(\protect\ref{eqtem}) to obtain the values of the divergent two-loop
three-point diagrams of Fig.~\protect\ref{fig1}.
$N=4$ for ${\cal F}_i$, $N=5$ for ${\cal G}_i$, and $N=6$ for ${\cal H}_i$.
Analytical and numerical results for ${\rm F}_2$, ${\rm f}_2$, ${\rm G}_5$,
${\rm g}_5$, ${\rm H}_5$, and ${\rm h}_5$ are given in
Eq.~(\protect\ref{eqcon}).
}
\label{tab1}
\medskip
\begin{tabular}{|c|c|c|l|} \hline\hline
Graph & $A$ &  \multicolumn{1}{c|}{ $B$ } & \multicolumn{1}{c|}{ $C$ }\\
\hline
${\cal F}_1 $ & $-\frac{1}{2}$ & $-\frac{5}{2}-i\pi$ & 
   $   \frac{11}{2}\zeta(2) -\frac{19}{2} - 5i\pi$ \\
${\cal F}_2 $ & $-\frac{1}{2}$ & $-\frac{5}{2}-i\pi$ & 
   $ {\rm F_2} + i \pi {\rm f_2} $\\
${\cal F}_3 $ & $-\frac{1}{2}$ & $\frac{\pi}{\sqrt3}-\frac{5}{2}$ & 
   $  \frac{8}{3}\zeta(3) -\frac{1}{3}\zeta(2) 
   + \left(-\frac{2\pi}{3}+\frac{1}{\sqrt3}\right)\cl
   \left(\frac{\pi}{3}\right) $ \\
& & & $ {}+ \frac{\pi}{\sqrt3}\left(-\ln3+5\right) -\frac{19}{2} $\\
${\cal F}_4 $ & $-\frac{1}{2}$ & $-\frac{3}{2}$ & 
   $ -\frac{5}{2}\zeta(2) -\frac{5}{2} $\\
${\cal F}_5 $ & $-\frac{1}{2}$ & $-\frac{3}{2}$ & 
   $  -\frac{5}{4}\zeta(3) +3\zeta(2)\left(\ln2 -\frac{1}{2} \right) 
     -\frac{5}{2} $\\
${\cal F}_6 $ & $-\frac{1}{2}$ & $-\frac{3}{2}$ &   
   $   2\zeta(3) -\frac{1}{2}\zeta(2) -\frac{5}{2} $\\
\hline
${\cal G}_1 $ & 0 & $ \frac{2}{3}\zeta(2) $ & 
   $ -\frac{11}{9}\zeta(3)
   +\frac{4}{3}\zeta(2)\left(-\frac{\pi}{3\sqrt3}+1\right)
   + \frac{4\pi}{9}\cl\left(\frac{\pi}{3}\right) $ \\
${\cal G}_2 $ & 0 & $ \frac{2}{3}\zeta(2) $ &
   $ \frac{1}{9}\zeta(3)+\frac{4}{3}\zeta(2)
   + \frac{4\pi}{9}\cl\left(\frac{\pi}{3}\right) $ \\
${\cal G}_3 $ & 0 & $ \frac{1}{2}\zeta(2) + i\pi\ln2 $ & 
   $ \frac{29}{8}\zeta(3)+\zeta(2)\left(-\frac{15}{2}\ln2+1\right) $ \\
& & & $ {}+ i\pi\left[\zeta(2)-\frac{1}{2}\ln^22+\ln2\right] $ \\
${\cal G}_4 $ & 0 & $ \frac{2}{3}\zeta(2) $ & 
   $ -\frac{26}{9}\zeta(3)-\frac{1}{6}\zeta(2)
   + \left(\frac{10\pi}{9}+\sqrt3\right)\cl\left(\frac{\pi}{3}\right)$\\
${\cal G}_5 $ & 0 & $ \frac{1}{2}\zeta(2) + i\pi\ln2 $ & 
   $ \frac{19}{8}\zeta(3)
   +\zeta(2)\left(-\frac{\pi}{2\sqrt3}-\frac{9}{2}\ln2+1\right)
   +{\rm G_5} $\\
& & & $ {}+ i\pi\left[\zeta(2)-\frac{1}{2}\ln^22 +
   \left(-\frac{\pi}{\sqrt3}+2\right)\ln2 + {\rm g_5}\right]  $\\
${\cal G}_6 $ & 0 & $ \frac{1}{2}\zeta(2) + i\pi\ln2 $ & 
   $ \frac{25}{8}\zeta(3)+\zeta(2)\left(-\frac{15}{2}\ln2+1\right) $ \\
& & & $ {}+ i\pi\left[\frac{3}{2}\zeta(2)-\frac{1}{2}\ln^22+2\ln2\right] $ \\
\hline
${\cal H}_1 $ & 0 & $ -\frac{\pi}{3\sqrt3} $ & 
   $ \zeta(2)\left(\frac{8\pi}{27\sqrt3}-\frac{3}{2}\right)
   + \frac{5}{3\sqrt3}\cl\left(\frac{\pi}{3}\right) 
   + \frac{\pi}{3\sqrt3}\left(\ln3 -2\right)$ \\
${\cal H}_2 $ & 0 & $ -\frac{\pi}{3\sqrt3} $ & 
   $ -\frac{5}{6}\zeta(2)
   - \frac{7}{3\sqrt3}\cl\left(\frac{\pi}{3}\right) 
   + \frac{\pi}{3\sqrt3}\left(\ln3 -2\right)$ \\
${\cal H}_3 - {\cal B}_1{\cal D}_1 $    & 0 & 0 & 
   $ \frac{3}{8}\zeta(2) -\frac{1}{2} + i \frac{\pi}{2}\ln2  $\\
${\cal H}_4 - {\cal B}_1{\cal D}_{2b} $ & 0 & 0 & 
   $ -\frac{1}{12}\zeta(2)  
   - \frac{\sqrt3}{2}\cl\left(\frac{\pi}{3}\right) 
   + \frac{\pi}{2\sqrt3}+\frac{1}{2}  $\\
${\cal H}_5 $ & 0 & $ -i\frac{\pi}{2} $ & 
   $ \zeta(2)\left(\frac{\pi}{3\sqrt3} + \frac{3}{4}\right) + {\rm H_5} $ \\
& & & $ {}+ i\pi\left[ \left(\frac{2\pi}{3\sqrt3}-\frac{5}{2}\right)\ln2 +
   \frac{\pi}{2\sqrt3} -1    + {\rm h_5} \right]  $\\
${\cal H}_6 $ & 0 & $ -i\frac{\pi}{2} $ & 
   $ \frac{9}{4}\zeta(2) - i \pi\left(\frac{5}{2}\ln2 + 1 \right) $\\
\hline\hline
\end{tabular}
\end{table}

\newpage 

\begin{table}[p]

\caption{Coefficients $C$ to be inserted in 
template~(\protect\ref{eqtem}) to obtain the values of the finite two-loop
three-point diagrams of Fig.~\protect\ref{fig1} ($A=B=0$).
$N=5$ for ${\cal I}_i$ and $N=6$ for ${\cal J}_i$ and ${\cal K}_i$.
The numerical uncertainties in the last digit(s) are given in parentheses.
}
\label{tab2}
\medskip
\begin{tabular}{|c|lclcl|} \hline\hline
Graph & \multicolumn{5}{c|}{ $C$ }\\
\hline
${\cal I}_1 $ &
  ${\rm I_1}$ & $=$ & $-1.894\,06(5)$ & & \\
${\cal I}_2 $ &
  ${\rm I_2}$ & $=$ & $-2.004\,830(5)$ & & \\
${\cal I}_3 $ &
  ${\rm I_3}$ & $=$ & $-3.652\,841(5)$ & & \\
${\cal I}_4 $ & 
  ${\rm I_4}$ & $=$ & $-1.280\,380(5)$ & & \\
${\cal I}_5 $ & 
  ${\rm I_5} + i\,{\rm i_5}$ & $=$
    & $-1.352\,904(5)$ & $-$ & $7.552\,73(3) \, i$ \\ 
${\cal I}_6 $ & 
  ${\rm I_6}$ & $=$ & $-1.921\,491(5)$ & & \\
${\cal I}_7 $ & 
  ${\rm I_7} + i\,{\rm i_7}$ & $=$ 
    & $-2.137\,588(5)$ & $-$ & $3.021\,09(3) \, i$\\ 
\hline
${\cal J}_1 $ & 
  ${\rm J_1}$ & $=$ & $\phantom{-}0.482\,636\,3(5)$ & & \\
${\cal J}_2 $ & 
 ${\rm J_2} + i\,{\rm j_2}$ & $=$ 
   & $\phantom{-}0.050\,486\,6(10)$ & $+$ & $2.388\,03(6) \, i$\\ 
${\cal J}_3 $ & 
 ${\rm J_3} + i\,{\rm j_3}$ & $=$ 
   & $-0.829\,866(5)$ & $+$ & $5.554\,50(3) \, i$\\ 
${\cal J}_4 $ & 
  ${\rm J_4} + i\,{\rm j_4}$ & $=$
    & $-3.890\,156(5)$ & $+$ & $1.675\,52(3) \, i$\\ 
${\cal J}_5 $ & 
  ${\rm J_5}$ & $=$ & $\phantom{-}0.779\,304(10)$ & & \\
${\cal J}_6 $ & 
  ${\rm J_6}$ & $=$ & $\phantom{-}2.067\,12(5)$ & & \\
${\cal J}_7 $ & 
  ${\rm J_7} + i\,{\rm j_7}$ & $=$
    & $\phantom{-}0.249\,480(10)$ & + & $2.103\,29(3) \, i$\\ 
${\cal J}_8 $ & 
  ${\rm J_8} + i\,{\rm j_8}$ & $=$
    & $-1.199\,53(5)$ & $+$ & $5.567\,4(2) \, i$\\ 
\hline
${\cal K}_1 $ & 
  ${\rm K_1}$ & $=$ & $\phantom{-}0.607\,011(5)$ & & \\
${\cal K}_2 $ & 
  ${\rm K_2} + i\,{\rm k_2}$ & $=$
    & $-1.211\,623(5)$ & $+$ & $4.995\,5(3) \, i$\\ 
\hline\hline
\end{tabular}
\end{table}

\newpage

\begin{figure}[p]
\vspace{0.05in}
\centerline{
  \epsfxsize=4.18in
  {\epsffile{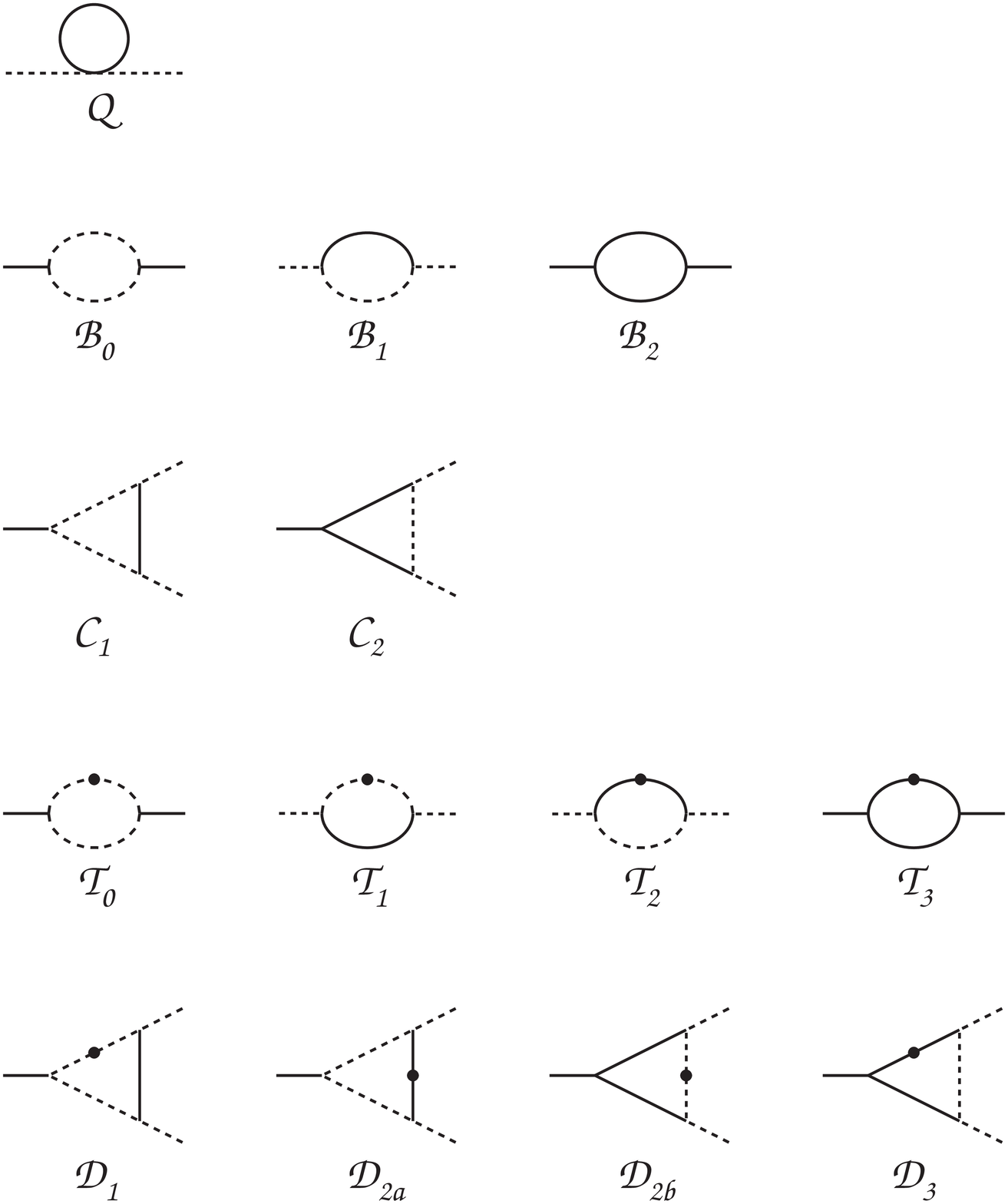}}
}
\vspace{0.05in}
\caption{Feynman diagrams pertinent to the $H\to VV$ 
transition amplitude through $O(G_F^2M_H^4)$.
Dashed (solid) lines represent Goldstone (Higgs) bosons.
Adjacent propagators with identical four-momenta are separated by a solid
circle.
}
\label{fig1}
\end{figure}

\newpage

\addtocounter{figure}{-1}
\begin{figure}[p]
\vspace{0.05in}
\centerline{
  \epsfxsize=3.04in
  {\epsffile{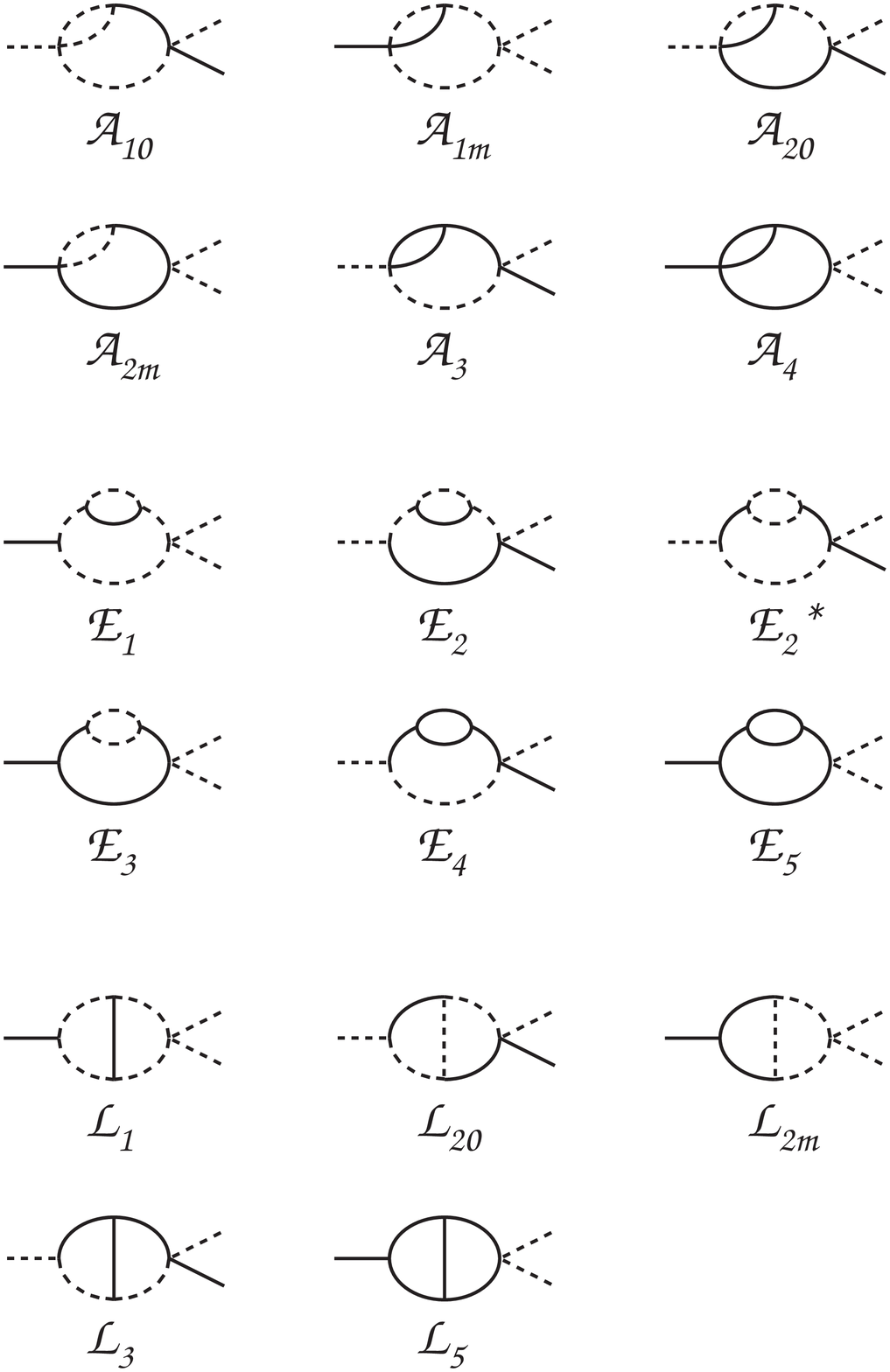}}
}
\vspace{0.05in}
\caption{Continued.}
\end{figure}

\newpage

\addtocounter{figure}{-1}
\begin{figure}[p]
\vspace{0.05in}
\centerline{
  \epsfxsize=3.04in
  {\epsffile{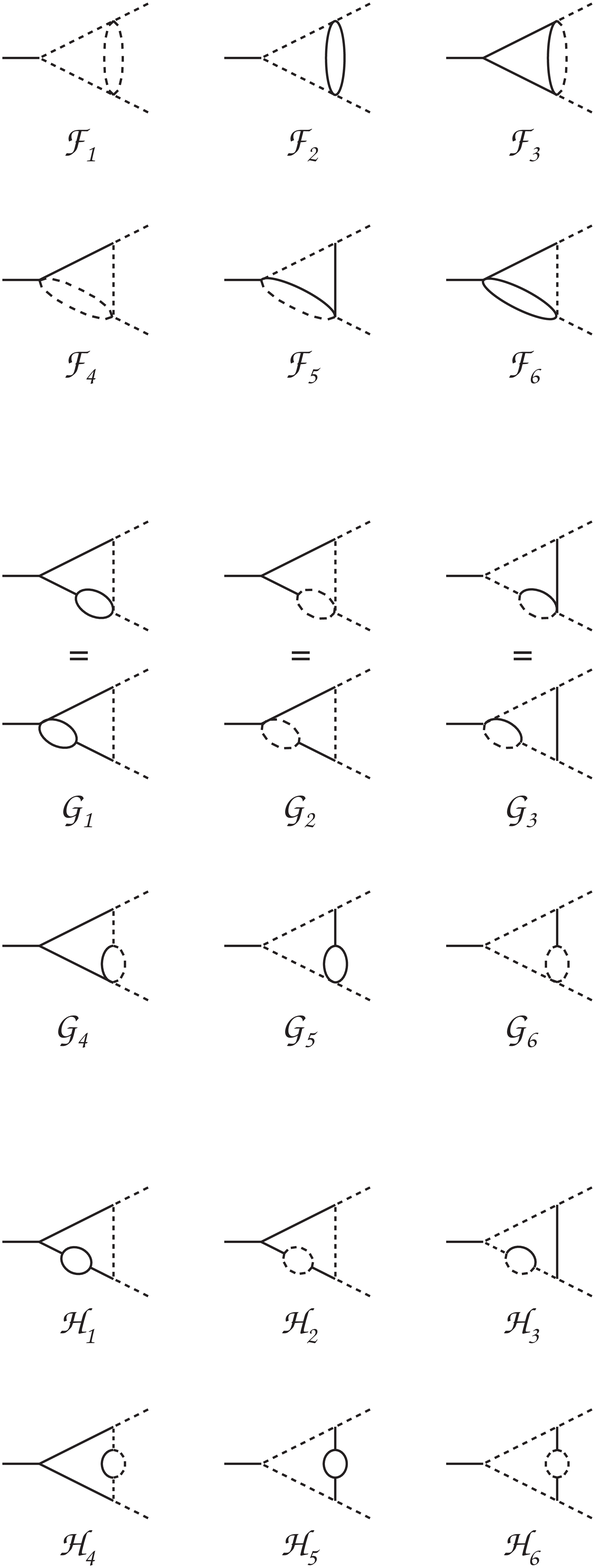}}
}
\vspace{0.05in}
\caption{Continued.}
\end{figure}

\newpage

\addtocounter{figure}{-1}
\begin{figure}[p]
\vspace{0.05in}
\centerline{
  \epsfxsize=4.18in
  {\epsffile{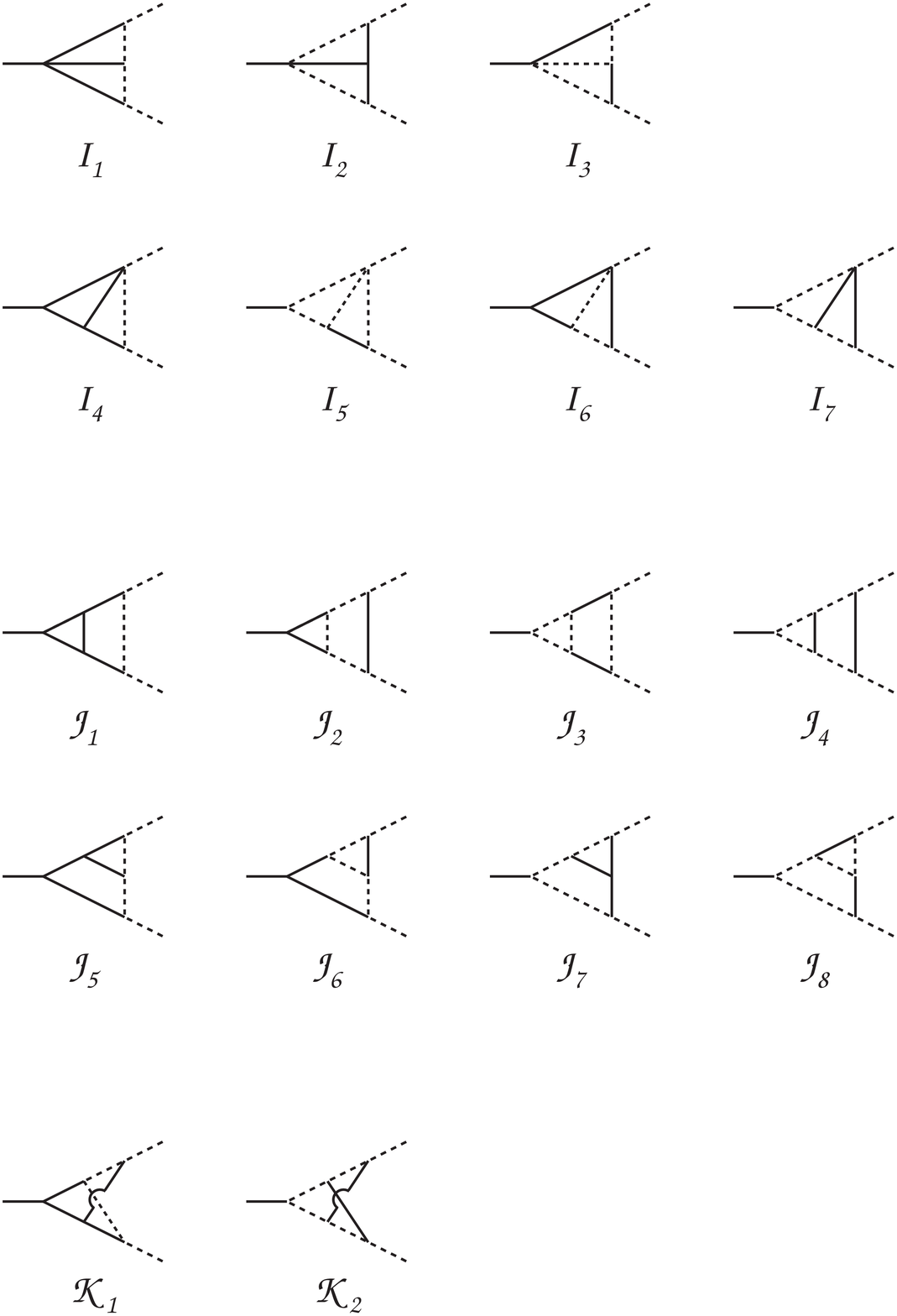}}
}
\vspace{0.05in}
\caption{Continued.}
\end{figure}

\newpage

\begin{figure}[p]
\vspace{0.05in}
\centerline{
  \epsfysize=4.0in
  \rotate[l]
  {\epsffile{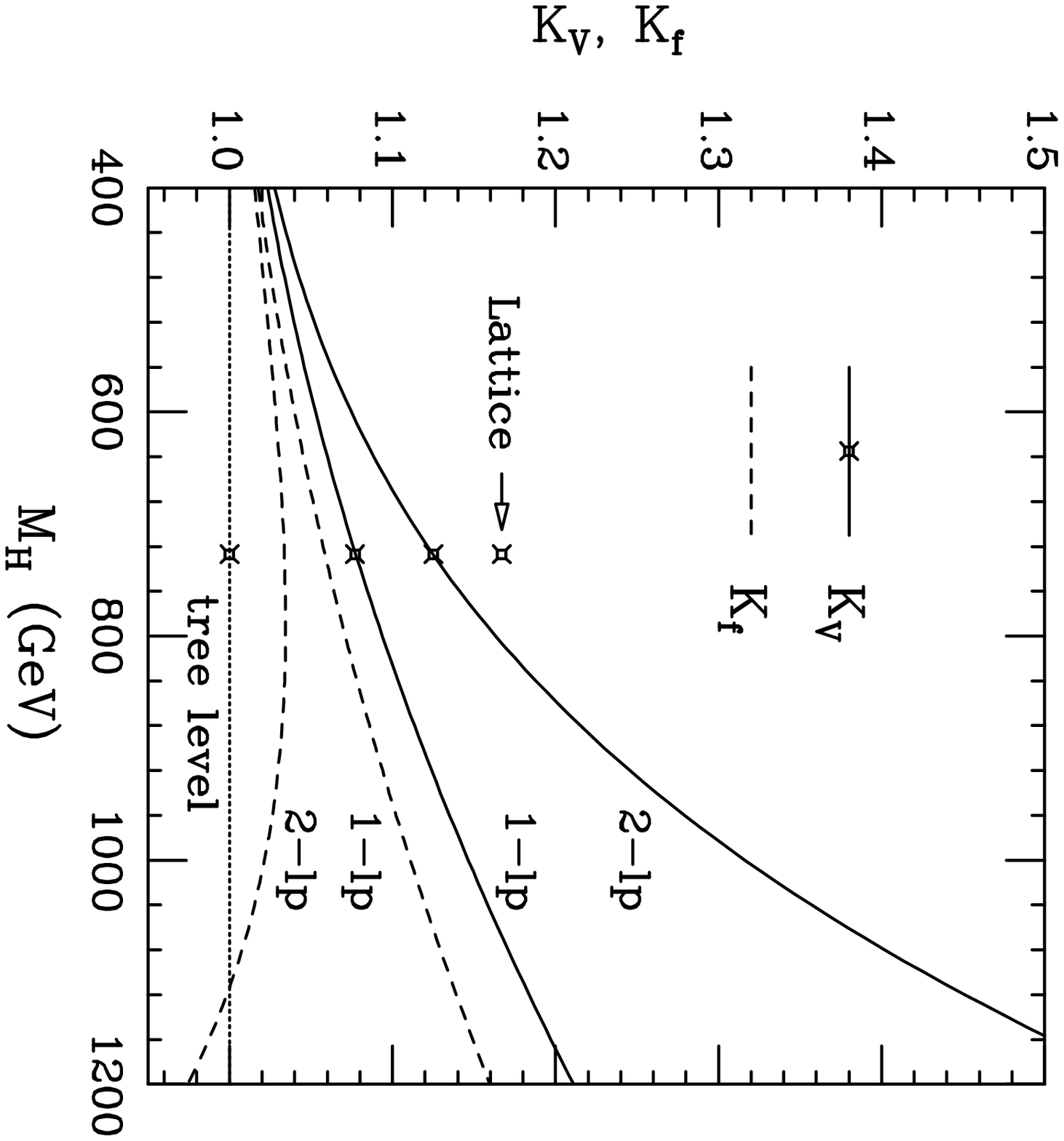}}
}
\vspace{0.05in}
\caption{$H\to VV$ correction factor $K_V$ of
Eq.~(\protect\ref{eqkv}) to $O(G_FM_H^2)$ and $O(G_F^2M_H^4)$ as a function of
$M_H$ (solid lines).
For comparison, we also show the $H\to f\bar f$ correction factor $K_f$ of
Eq.~(\protect\ref{eqkf}) to $O(G_FM_H^2)$ and $O(G_F^2M_H^4)$ (dashed lines).
The crosses indicate the tree-level, one-loop, two-loop, and nonperturbative
values of $K_V$ at $M_H=727$~GeV.
}
\label{fig2}
\end{figure}

\newpage

\begin{figure}[p]
\vspace{0.05in}
\centerline{
  \epsfysize=4.0in
  \rotate[l]
  {\epsffile{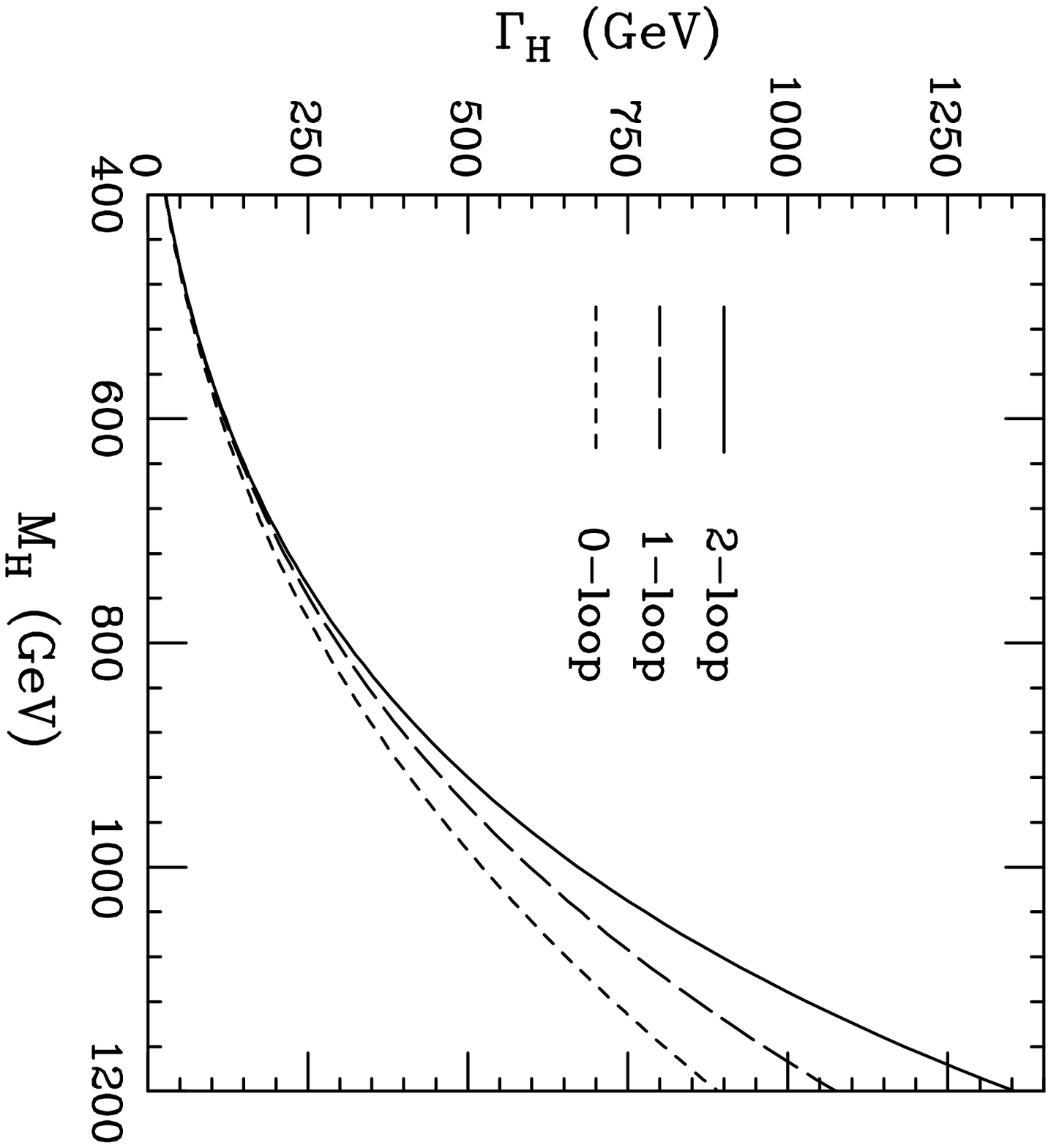}}
}
\vspace{0.05in}
\caption{Total Higgs-boson decay width $\Gamma_H$ in the
Born approximation (short-dashed line), to $O(G_FM_H^2)$ (long-dashed line),
and to $O(G_F^2M_H^4)$ (solid line) as a function of $M_H$.
We assume $M_t=175$~GeV.
}
\label{fig3}
\end{figure}

\end{document}